\documentclass[preprint,12pt]{elsarticle}

\usepackage{amssymb}
 \usepackage{amsmath}
\usepackage{amsmath,bm}
\usepackage{color}

\journal{Advances in High Energy Physics}

\begin{document}

\begin{frontmatter}



\title{Influence of Lorentz Invariation
Violation on Arbitrarily Spin Fermions Tunneling Radiation in the
Vaidya-Bonner Spacetime}

 \author[label1,label2]{Jie Zhang}

\author[label1]{Zhie Liu}
\author[label1]{Bei Sha}
\author[label1]{Xia Tan}
\author[label1]{Yuzhen Liu}
\author[label2]{Shuzheng Yang}

\address[label1]{College of Physics and Electronic Engineering, Qilu Normal University, Jinan  250300, China}
\address[label2]{College of Physics and Space science, China West  Normal University, Nanchong  637009,
China}

\begin{abstract}
In the spacetime of non-stationary spherical symmetry Vaidya-Bonner
black hole, an accurate modification of Hawking tunneling radiation
for fermions with arbitrarily spin is researched. Considering a
light dispersion relationship derived from string theory, quantum
gravitational theory and Rarita-Schwinger Equation in the
non-stationary spherical symmetry spacetime, we derive an accurately
modified dynamic equation for fermions with arbitrarily spin. By
solving the equation, modified tunneling rate of fermions with
arbitrarily spin, Hawking temperature and entropy at the event
horizon of Vaidya-Bonner black hole are presented. We find the
Hawking temperature will increase, but the the entropy will decrease
comparing with the case without Lorentz Invariation Violation
modification.
\end{abstract}



\end{frontmatter}

\section{Introduction}\label{section1}
The theory of Hawking thermal radiation reveals the relationship
between gravitational theory, quantum theory and statistical thermal
dynamic mechanics\cite{haw74}. After the research of Hawking thermal
radiation to all kinds of black holes\cite{yan95}, Kraus and Wilczek
did some modification to  Hawking thermal radiation adopting
self-gravitational interaction\cite{kra95}. Hereafter, researchers
studied the Hawking tunneling radiation for many types of black
holes \cite{par00,zha05,yan05,ovg16,kua17,sak17}. In 2007, Kerner
and Mann et al. proposed a semi-classic method to investigate the
tunneling radiation of fermions with spin
$\frac{1}{2}$\cite{ker08a,ker08b}. In the later research, this
semi-classic method is widely used to calculate the tunneling
radiation of the other type of particles\cite{yal09,lin09a,cri08}.
Yang and Lin developed Kerner and Mann's theory and proposed that
Hamilton-Jacobi method is efficient to the tunneling of
fermions\cite{lin09b,yan10}. According to
references\cite{lin09b,yan10}, after choosing suitable Gamma matrix
and considering commutation relation of the Pauli matrix in Dirac
Equation, which describes the dynamic of fermion quite well,
Hamilton-Jacobi Equation in the curved spacetime can be derived.
This result means that Hamilton-Jacobi Equation is also a very
important equation in the research of tunneling theory of fermion.
In recent years, Lorentz light dispersion relationship is generally
regarded as a basic relation in the modern physics. It seems that
both general relativity and quantum mechanics are built on this
relationship. However, the research of quantum gravitational theory
indicates the Lorentz relationship should be modified in the high
energy case. Although scientists have not built a successful light
dispersion relationship in the high energy  case, current researches
are helpful to the development of this theory. People usually
estimate the magnitude of this modification should be in Plank
scale. It is confirmed that both Dirac Equation and Hamilton-Jacobi
Equation must be modified if the Lorentz Invariation Violation is
considered. In such case, only accurate modification can efficiently
research fermion tunneling radiation from black hole, such as
Vaidya-Bonner black hole. In this paper, the most important progress
is that we use a new method which is suitable for fermions with
arbitrarily spin. We will research the exact modification of
tunneling radiation for fermions with arbitrarily spin, considering
the Lorentz Invariation Violation.

\section{Exact Modification of Arbitrarily Spin Fermions Rarita-Schwinger Equation and Hamilton-Jacobi Equation}\label{sec:2}
 In the research of string theory, the authors proposed a
 relation\cite{ame02,mag02,ell92,kru13,jac03}
 \begin{equation} \label{eq1}
P_0^2=p^2+m^2-(LP_0)^\alpha p^2.
  \end{equation}
In the natural unit, $P_0$, $p$ are the energy and momentum of
particle with the static mass $m$. $L$ is a constant in the
magnitude of Plank scale, which comes from Lorentz Invariation
Violation theory. In Eq.(\ref{eq1}), $\alpha=1$ is adopted in the
Liouville-string model. Kruglov obtained a modified Dirac equation
considering $\alpha=2$\cite{kru12}. Therefore, we substitute
$\alpha=2$ into Eq.(\ref{eq1})  and get a general Rarita-Schwinger
equation in the flat space
\begin{equation} \label{eq:2}
[\bar{\gamma}^{\mu}\partial_\mu+\frac{m}{\hbar}-\lambda \hbar
\bar{\gamma}^t\partial_t\bar{\gamma}^j\partial_j]\Psi_{\alpha_1\ldots\alpha_k}=0,
 \end{equation}
where $\hbar$ is the reduced Plank constant, which equals 1 in the
natural units.  $\lambda$ is a very small constant.
$\Psi_{\alpha_1\ldots\alpha_k}$ is wave function, where the value of
$\alpha_k$ corresponds to different spin. The larger $\alpha_k$, the
higher the spin is. The wave function satisfies following
supplementary condition:
\begin{equation} \label{eq:3}
\bar{\gamma}^\mu\Psi^\mu_{\alpha_2\ldots\alpha_k}=\partial_\mu
\Psi^\mu_{\alpha_2\ldots\alpha_k}=\Psi^\mu_{\mu_{\alpha_3\ldots\alpha_k}}=0.
 \end{equation}
 When $k=0$, $\Psi_{\alpha_1\ldots\alpha_k}=\Psi$, Eq.(\ref{eq:2})
 changes to the Dirac equation for spin $\frac{1}{2}$ and condition
 (\ref{eq:3}) disappears automatically. When $k=1$, Eq.(\ref{eq:2})
 describes the dynamic of fermions with spin
 $\frac{3}{2}$ and the condition (\ref{eq:3}) also disappears automatically. Note that the commutation relation
 \begin{equation} \label{eq:4}
\{\gamma^{\mu},\gamma^{\nu}\}=2g^{\mu \nu}I.
  \end{equation}
In the curved spacetime, Rarita-Schwinger equation can be rewritten
as
\begin{equation} \label{eq:6}
[\gamma ^{\mu}D_\mu+\frac{m}{\hbar}-\lambda \hbar \gamma^t
D_t\gamma^j D_j]\Psi_{\alpha_1\ldots\alpha_k}=0,
 \end{equation}
 where $D_{\mu}=\partial_{\mu}+\frac{{\rm i}}{\hbar} e A_{\mu}$, $\lambda<<1$, $\lambda \hbar \gamma^t
D_t\gamma^j D_j$ is a very small term. For fermions with arbitrarily
spin, the wave function is
\begin{equation} \label{eq:7}
\Psi_{\alpha_1\ldots\alpha_k}=\xi_{\alpha_1\ldots\alpha_k}e^{\frac{{\rm
i}}{\hbar}S},
 \end{equation}
 where $\xi_{\alpha_1\ldots\alpha_k}$ and $S$ are matrices and the action of fermion, respectively.
The line element of the non-stationary Vaidya-Bonner black hole
represented in advanced Eddington coordinate \cite{bon70} is given
by
\begin{equation} \label{eq:8}
ds^2=-F(r,v)dv^2+2drdv+r^2(d\theta^2+\sin^2\theta d \varphi^2),
 \end{equation}
\begin{equation} \label{eq:9}
 F(r,v)=1-\frac{2M(v)}{r}+\frac{Q^2(v)}{r^2},
 \end{equation}
 where $v$  is the Eddington time,   $M(v)$ and $Q(v)$ represent
the mass and charge of the black hole changes with time
respectively. When $Q(v) = 0$, the non-stationary Vaidya-Bonner
black hole is reduced to the Vaidya black hole. The electromagnetic
four potential of Vaidya-Bonner black hole is
\begin{equation} \label{eq:10}
A_\mu=(\frac{Q}{r},0,0,0)=(A_0,0,0,0).
 \end{equation}
Corresponding to line element (\ref{eq:8}), the inverse metric
tensor is
\begin{equation} \label{eq:11}
g^{\mu\nu}= \left( \begin{array}{cccc} 0& 1&0&0 \\ 1&
\frac{\Delta}{r^2}&0&0\\0&0&r^{-2}&0\\0&0&0&r^{-2}\sin^{-2}\theta\end{array}
\right),
 \end{equation}
 where
 \begin{equation} \label{eq:12}
\Delta=r^2-2Mr+Q^2.
  \end{equation}
  Because the component of inverse metric tensor $g^{00}=0$ in the curved
  spacetime of line element (\ref{eq:8}),  so Eq.(\ref{eq:6}) and Eq.(\ref{eq:7}) becomes
  \begin{equation} \label{eq:13}
[{\rm i} \gamma^0(\partial _{v}S+eA_0)+{\rm i} \gamma^j\partial
_{j}S+m+\lambda
(\partial_{v}S+eA_0)\gamma^0\gamma^i\partial_iS]\xi_{\alpha_1\ldots\alpha_k}=0.
   \end{equation}
In this paper, the range for $i$ and $j$ in the superscript and
subscript satisfies $i,j=1,\ 2,\ 3$.  $\mu$ and $\nu$ in the
superscript and subscript are defined as $\mu,\nu=0,\ 1,\ 2,\ 3$.
Setting
\begin{equation} \label{eq:14}
\Gamma^{\mu}={\rm
i}\gamma^{\mu}+\lambda(\partial_{v}S+eA_0)\gamma^0\gamma^\mu,
 \end{equation}
Eq.(\ref{eq:13}) becomes
\begin{equation} \label{eq:15}
[m+\Gamma^{\mu}
(\partial_{\mu}S+eA_{\mu})]\xi_{\alpha_1\ldots\alpha_k}=0.
 \end{equation}
Multiplying $\Gamma^{\nu} (\partial_{\nu}S+eA_{\nu})$ in the both
sides of Eq.(\ref{eq:15}), then
\begin{equation} \label{eq:16}
\Gamma^{\nu} (\partial_{\nu}S+eA_{\nu})\Gamma^{\mu}
(\partial_{\mu}S+eA_{\mu})\xi_{\alpha_1\ldots\alpha_k}-m^2\xi_{\alpha_1\ldots\alpha_k}=0.
 \end{equation}
Exchanging $\nu$ and $\mu$ in Eq.(\ref{eq:16}), we get
\begin{equation} \label{eq:17}
\Gamma^{\mu} (\partial_{\mu}S+eA_{\mu})\Gamma^{\nu}
(\partial_{\nu}S+eA_{\nu})\xi_{\alpha_1\ldots\alpha_k}-m^2\xi_{\alpha_1\ldots\alpha_k}=0.
 \end{equation}
 Eq.(\ref{eq:16}) and Eq.(\ref{eq:17}) are equivalent. Considering
 $\gamma^0\gamma^0=g^{00}=0$,  firstly adding the left side and right side of Eq.(\ref{eq:16}) and
 Eq.(\ref{eq:17}) respectively, and then dividing  the new equation by 2, finally combing with Eq.(\ref{eq:14}), we obtain
 \begin{equation} \label{eq:18}
  \begin{split}\{2g^{0j}(\partial_{v}S+eA_{0})\partial_{j}S+g^{ij}\partial_{i}S\partial_{j}S\\-{\rm i}2\lambda(\partial_{v}S+eA_{0})g^{0i}\partial_{i}S\gamma^{\mu}
(\partial_{\mu}S+eA_{\mu})
\\-\lambda^2[(\partial_{v}S+eA_{0})g^{0j}\partial_{j}S]^2+m^2\}\xi_{\alpha_1\ldots\alpha_k}=0.
\end{split}
  \end{equation}
Defining
\begin{equation} \label{eq:19}
\begin{split}
m_l=\frac{-2g^{0j}(\partial_{v}S+eA_{0})\partial_{j}S-g^{ij}\partial_{i}S\partial_{j}S}
{2(\partial_{v}S+eA_{0})g^{0i}\partial_{i}S}+\\
\frac{-m^2+\lambda^2[(\partial_{v}S+eA_{0})g^{0j}\partial_{j}S]^2}
{2(\partial_{v}S+eA_{0})g^{0i}\partial_{i}S},
 \end{split}\end{equation}
Eq.(\ref{eq:18}) changes to
\begin{equation} \label{eq:20}
{\rm i}\lambda
\gamma^{\mu}(\partial_{\mu}S+eA_{\mu})\xi_{\alpha_1\ldots\alpha_k}+m_l\xi_{\alpha_1\ldots\alpha_k}=0.
 \end{equation}
Multiplying ${\rm i}\lambda \gamma^{\nu}(\partial_{\nu}S+eA_{\nu})$
at both sides, we get
\begin{equation} \label{eq:21}
\lambda^2
\gamma^{\mu}\gamma^{\nu}(\partial_{\mu}S+eA_{\mu})(\partial_{\nu}S+eA_{\nu})\xi_{\alpha_1\ldots\alpha_k}+m^2_l\xi_{\alpha_1\ldots\alpha_k}=0.
 \end{equation}
By exchanging $\mu$ and $\nu$ for Eq.(\ref{eq:21}), then
\begin{equation} \label{eq:22}
\lambda^2
\gamma^{\nu}\gamma^{\mu}(\partial_{\nu}S+eA_{\nu})(\partial_{\mu}S+eA_{\mu})\xi_{\alpha_1\ldots\alpha_k}+m^2_l\xi_{\alpha_1\ldots\alpha_k}=0.
 \end{equation}
Combining Eq.(\ref{eq:21}), Eq.(\ref{eq:22}) and
\begin{equation} \label{eq:23}
\gamma^{\nu}\gamma^{\mu}+\gamma^{\mu}\gamma^{\nu}=2g^{\mu \nu}I,
 \end{equation}
it is easy to get
\begin{equation} \label{eq:24}
[\lambda^2g^{\mu \nu}
(\partial_{\mu}S+eA_{\mu})(\partial_{\nu}S+eA_{\nu})+m^2_l]\xi_{\alpha_1\ldots\alpha_k}=0.
 \end{equation}
Eq.(\ref{eq:24}) is a matrix equation. In fact, it is an eigenvalue
matrix equation. The condition for nonsingular solution of this
eigenvalue matrix equation requires the corresponding value of
determinant is zero. Combining Eq.(\ref{eq:19}), Eq.(\ref{eq:24})
and Eq.(\ref{eq:8}), we get
\begin{equation} \label{eq:25}
\begin{split}
\frac{2g^{0j}(\partial_{v}S+eA_{0})\partial_{j}S+g^{ij}\partial_{i}S\partial_{j}S+m^2}
{2(\partial_{v}S+eA_{0})g^{0i}\partial_{i}S}+\\
\frac{-\lambda^2[(\partial_{v}S+eA_{0})g^{0j}\partial_{j}S]^2}
{2(\partial_{v}S+eA_{0})g^{0i}\partial_{i}S} -\lambda m=0.
 \end{split}\end{equation}
Therefore,
\begin{equation} \label{eq:26}
\begin{array}{c}
2g^{0j}(\partial_{v}S+eA_{0})\partial_{j}S+g^{ij}\partial_{i}S\partial_{j}S+m^2-2
\lambda m(\partial_{v}S+eA_{0})g^{0i}\partial_{i}S\\
-\lambda^2[(\partial_{v}S+eA_{0})g^{0j}\partial_{j}S]^2=0.
\end{array}
 \end{equation}
As $\lambda<<1$, $o(\lambda^2)$ is a high order term. For accuracy
of modification, the term $o(\lambda^2)$ is kept in
Eq.(\ref{eq:18}), Eq.(\ref{eq:24}) and  Eq.(\ref{eq:25}). If
$o(\lambda^2)$ is ignored in Eq.(\ref{eq:18}), one cannot obtain a
correct result. For Vaidya-Bonner black hole, only this derivation
can get a correct result. In fact, Eq.(\ref{eq:26}) is the dynamic
equation describing an arbitrarily spin fermion in the Vaidya-Bonner
spacetime. Moreover, this equation is derived from Rarita-Schwinger
Equation in the curved spacetime with Lorentz Invariance Violation.
So this equation is a deformation of Hamilton-Jacobi Equation, or
exactly can be called as Rarita-Schwinger-Hamilton-Jacobi Equation.
The first two terms of this equation can be expressed as
$g^{\mu\nu}(\partial_{\mu}S+eA_{\mu})(\partial_{\nu}S+eA_{\nu})$.
Considering Eq.(\ref{eq:11}), one can obtain the first two terms in
Eq.(\ref{eq:26}), so Eq.(\ref{eq:26}) can be rewritten as
\begin{equation} \label{eq:27}
\begin{array}{c}
g^{\mu\nu}(\partial_{\mu}S+eA_{\mu})(\partial_{\nu}S+eA_{\nu})+m^2-2
\lambda m(\partial_{v}S+eA_{0})g^{0i}\partial_{i}S\\
-\lambda^2[(\partial_{v}S+eA_{0})g^{0j}\partial_{j}S]^2=0.
\end{array}
 \end{equation}
 From this equation, we can get the action of fermion, and then
 study the modified tunneling radiation of fermions. Eq.(\ref{eq:27}) is a highly accurate dynamic equation
 because the term $o(\lambda^2)$ is not ignored during the derivation and Lorentz Invariance Violation is
 included.
 If not do so, an accurate modification cannot be obtained. Note that the modification of Bosons's Hamilton-Jacobi
 Equation is different from this method \cite{yan16,fen19}. This indicates
 the significance of accurate modification of tunneling for particles with arbitrarily
 spin. In the following we will derive the thermal dynamic
 characteristics at the horizon of the Vaidya-Bonner black hole.

\section{Tunneling Modification  for Fermions with Arbitrarily
 Spin in Vaidya-Bonner Black Hole}\label{sec:3}
 Vaidya-Bonner black hole is a charged non-stationary spherical black hole,
  the line element has been shown in Eq.({\ref{eq:8}}). The event horizon of
 Vaidya-Bonner black hole is determined by the zero supercurved equation
\begin{equation} \label{eq:28}
g^{\mu\nu}\frac{\partial f}{\partial x^{\mu}}\frac{\partial
f}{\partial x^{\nu}}=0.
 \end{equation}
From  Eq.({\ref{eq:11}}) and  Eq.({\ref{eq:28}}), we find the event
horizon $r_H$ satisfies
\begin{equation} \label{eq:29}
r^2_H-2Mr_H+Q^2-2\dot{r}_Hr^2_H=0,
 \end{equation}
where $\dot{r_H}=\frac{dr_H}{d v}$ is the change rate of $r_H$ with
time. Solving Eq.(\ref{eq:29}) we get
\begin{equation} \label{eq:30}
r_H=\frac{M\pm[M^2-Q^2(1-2\dot{r}_H)]^{1/2}}{1-2\dot{r}_H},
 \end{equation}
where $'+'$ denotes the event horizon of the Vaidya-Bonner black
hole. From Eq.(\ref{eq:11}) and Eq.(\ref{eq:26}), the accurate
dynamic equation of  fermions with arbitrarily spin is
\begin{equation} \label{eq:31}
\begin{split}
\frac{\Delta}{r^2}(\frac{\partial S}{\partial
r})^2+\frac{1}{r^2}(\frac{\partial S}{\partial
\theta})^2+\frac{1}{r^2\sin^2\theta}(\frac{\partial S}{\partial
\varphi})^2+2(\frac{\partial S}{\partial v}+eA_{0})(\frac{\partial S}{\partial r})\\
+m^2-2 \lambda m(\frac{\partial S}{\partial
v}+eA_{0})(\frac{\partial S}{\partial r})-\lambda^2(\frac{\partial
S}{\partial v}+eA_{0})^2(\frac{\partial S}{\partial r})^2=0.
\end{split}
 \end{equation}
The key to research the tunneling is to get the action $S$ of
fermions. For this black hole, the key is solution of the action in
the direction of radius, so a tortoise coordinate transformation is
necessary,
\begin{equation} \label{eq:32}
r_{*}=r+\frac{1}{2\kappa}\ln\frac{r-r_H(v)}{r_H(v_0)},
 \end{equation}
\begin{equation} \label{eq:33}
v_{*}=v-v_{0},
 \end{equation}
where $\kappa$ is the surface gravity, $r_H$ is the event horizon of
the black hole, $v_0$ is a special moment when fermion escapes from
the event horizon. Both $v_0$ and $\kappa$ are constants. From
Eq.(\ref{eq:32}) and Eq.(\ref{eq:33}), we have
\begin{equation} \label{eq:34}
\frac{\partial}{\partial r}=[1+\frac{1}{2\kappa
(r-r_H)}]\frac{\partial}{\partial r_{*}},
 \end{equation}
 \begin{equation} \label{eq:35}
\frac{\partial}{\partial v}=[\frac{\partial}{\partial
v_{*}}+\frac{\dot{r}_H}{2\kappa (r-r_H)}]\frac{\partial}{\partial
r_{*}}.
  \end{equation}
We make a separation of variable to action $S$ as
\begin{equation} \label{eq:36}
S(v_{*},r_H,\theta,\varphi)=R(v_{*},r_H)+Y(\theta,\varphi),
 \end{equation}
and set
\begin{equation} \label{eq:37}
\frac{\partial S}{\partial v_{*}}=\frac{\partial R}{\partial
v_{*}}=-\omega.
 \end{equation}
Substituting Eq.(\ref{eq:32}-\ref{eq:37}) into
 Eq.(\ref{eq:31}), then Eq.(\ref{eq:31}) becomes
\begin{equation} \label{eq:38}
\begin{array}{c}
(\frac{\partial{R}}{\partial
r_{*}})^2\{\frac{\Delta}{r^2}[\frac{1+2\kappa(r-r_H)}{2 \kappa
(r-r_H)}]^2+\frac{2\dot{r}_H[1+2\kappa(r-r_H)]}{[2\kappa(r-r_H)]^2}-2\lambda
m\dot
r_H\frac{[1+2\kappa(r-r_H)]}{[2\kappa(r-r_H)]^2}\\
-\lambda^2\frac{1}{[2\kappa(r-r_H)]^2} (\frac{\partial{R}}{\partial
v_{*}}+eA_0)^2[1+2\kappa(r-r_H)]\}\\
+2(\frac{\partial{R}}{\partial v_{*}}+eA_0)(1-\lambda
m)\frac{[1+2\kappa(r-r_H)]}{\kappa(r-r_H)}\frac{\partial R}{\partial
r_{*}}+m^2+\frac{\lambda}{r^2}+o(\lambda^{'2})=0,
\end{array}
 \end{equation}
where  $\lambda^{'}=\lambda\dot r_H$. For simplification, it is
suitable to  keep the first-order term of $\lambda$ in the final
results. Multiplying $2\kappa(r-r_H)$ at the both sides of
Eq.(\ref{eq:38}), and taking limit for the condition $r\rightarrow
r_H$, we get
\begin{equation} \label{eq:39}
(\frac{\partial{R}}{\partial r_{*}})^2-2(1-\lambda
m)(\omega-\omega_0)(\frac{\partial{R}}{\partial r_{*}})=0,
 \end{equation}
 where $\omega_0=\frac{eQ}{r_H}$. The limit of the coefficient of $(\frac{\partial{R}}{\partial
r_{*}})^2$ is
\begin{equation} \label{eq:40}
\begin{split}
\lim_{r\rightarrow r_H} \frac{1}{2\kappa(r-r_H)r^2}[(r^2-2Mr+Q^2)[1+2\kappa(r-r_H)]+\\
2\dot{r}_Hr^2-2\lambda m\dot r_H
r^2-\lambda^2(\omega-\omega_0)^2r^2]=1.
\end{split}
 \end{equation}
From Eq.(\ref{eq:34}) and Eq.(\ref{eq:39})
\begin{equation} \label{eq:42}
\begin{split}
\frac{\partial R}{\partial
r}&=\frac{1+2\kappa(r-r_H)}{2\kappa(r-r_H)}\frac{\partial
R}{\partial
r_{*}}\\
&=2\frac{1+2\kappa(r-r_H)}{2\kappa(r-r_H)}(1-\lambda
m)(\omega-\omega_0).
 \end{split}\end{equation}
Using the residue theorem to solve $R$ in Eq.(\ref{eq:42}), we get
\begin{equation} \label{eq:43}
\begin{split}
R_{\pm}&=\int{2\frac{1+2\kappa(r-r_H)}{2\kappa(r-r_H)}(1-\lambda
m)(\omega-\omega_0)dr}\\
&=\frac{{\rm i}\pi}{\kappa}(1-\lambda
m)[(\omega-\omega_0)\pm(\omega-\omega_0)].
 \end{split}\end{equation}
 From Eq.(\ref{eq:40}), $\kappa$ in Eq.(\ref{eq:43}) can be
 gotten as
 \begin{equation} \label{eq:44}
\kappa=\frac{r_H-M-2\dot r_H r_H-2\lambda m \dot r_H r_H-\lambda
^2(\omega-\omega_0)^2r_H}{2Mr_H-Q^2}.
  \end{equation}
Due to $\lambda\ll 1$, final result can only remain $\lambda$ term.
In Eq.(\ref{eq:43}), "+" and "-" represent outgoing and ingoing wave
from the horizon of the back hole. So according
 to the tunneling theory, we get the tunneling rate for fermions with arbitrarily
 spin in the Vaidya-Bonner spacetime.
 \begin{equation} \label{eq:45}
\begin{split}
\Gamma &=\exp(-2{\rm Im} S ) \\&=\exp(-2 {\rm Im} R_{\pm})\\
&=\exp[-\frac{2\pi}{\kappa^{'}}(\omega-\omega_0)]\\
&=\exp(-\frac{\omega-\omega_0}{T^{'}_H}).
   \end{split}\end{equation}
The $\kappa^{'}=\frac{\kappa}{1-\lambda m}$ in Eq.(\ref{eq:44}) and
Eq.(\ref{eq:45}) is the modified surface gravitational force at the
event horizon of the black hole. $T_H$ in Eq.(\ref{eq:45}) is
\begin{equation} \label{eq:46}
\begin{split}
 T_H^{'}&=\frac{\kappa^{'}}{2\pi}\\
 &=\frac{r_H-M-2\dot r_H
r_H-2\lambda m \dot r_H
r_H-\lambda^2(\omega-\omega_0)^2r_H}{2\pi(1-\lambda m)(2Mr_H-Q^2)}.
\end{split}
 \end{equation}

 This is a new form of Hawking temperature after modification
in the Vaidya-Bonner black hole. Obviously, the tunneling rate and
temperature of black hole have been significantly modified. Adopting
Taylor expansion for $1/(1-\lambda m)$  and neglecting the high
order items of $\lambda$, Eq.(\ref{eq:46}) becomes
\begin{equation} \label{eq:46b}
\begin{split}
 T_H^{'}&=\frac{r_H-M-2\dot r_H
r_H-2\lambda m \dot r_H
r_H-\lambda^2(\omega-\omega_0)^2r_H}{2\pi(2Mr_H-Q^2)}[1+\lambda
m+(\lambda
m)^2+\cdots]\\
&\approx\frac{r_H-M-2\dot r_H r_H+\lambda m( r_H-M-4\dot r_H
r_H)}{2\pi(2Mr_H-Q^2)} \\&=T_H+\frac{\lambda m( r_H-M-4\dot r_H
r_H)}{2\pi(2Mr_H-Q^2)},
\end{split}
 \end{equation}
where $T_H$ is the Hawking temperature without Lorentz Invariation
Violation modification. Usually, $\dot{r}_H\ll r_H$, so the Hawking
temperature becomes higher than that without Lorentz Invariation
Violation modification.

\section{Conclusions and Discussions}\label{sec:4}
In this paper, based on the modified Dirac equation proposed by
Krugov, we extend his work to the Rarita-Schwinger equation which
can describe fermions with arbitrarily spin, and accurately modify
the semi-classic Hamilton-Jacobi Equation. The characteristics of
tunneling radiation from a non-stationary spherical Vaidya-Bonner
black hole is derived. The results show that the tunneling rate,
 surface gravitational force and Hawking temperature, all of them should be
modified by a term related to parameter $\lambda$. Although it is a
minor modification term, it is still valuable for further research.
The tunneling rate and Hawking temperature indicate all these
characteristics  are still spherical symmetry. The another important
parameter in thermal kinetic is entropy. The modification of Hawking
temperature must induce the change of entropy. According to the
first law of thermal kinetic of black hole,
\begin{equation} \label{eq:48}
dM=TdS +VdJ +UdQ,
 \end{equation}
where $V$ and $U$ are the rotation potential and electromagnetic
potential of the black hole respectively.  For Vaidya-Bonner black
hole, the modified entropy at $r=r_H$ is
\begin{equation} \label{eq:49}
dS=\frac{dM  -UdQ}{T^{'}_H}.
 \end{equation}
  Eqs.(\ref{eq:45})
-(\ref{eq:49}) show a few new results. We find the
 modification of Hawking temperature and entropy is not only related
 to $\lambda$ but also related to $\dot{r}_H$ for the Vaidy-Bonner
 black hole. For the general case of $\dot{r}_H\ll r_H$, the Hawking temperature will increase, but
the the entropy will decrease comparing with the fiducial results
without Lorentz Invariation Violation modification. As
$\dot{r}_H=0$, our results can reduce to  the reults of
Reissner-Nordstr$\ddot{\rm o}$m black hole; as $\dot{r}_H=Q=0$, our
results can reduce to the results of Schwarchild black hole.

 For the curved spacetime with a component of
 inverse  metric tensor $g^{00}=0$, the modified results will include $\lambda$ and $\lambda^2$. In the results, keeping the main
 $\lambda$ term is enough since $\lambda$ is very small.
     In our work, we have chosen the condition $\alpha=2$ for the modified light
     dispersion relation. In general case, this condition should
     be cut off. Therefore, a more general modification method to
     Rarita-Schwinger equation and the tunneling it describes
     should be further discussed.

    Moreover, for the tunneling radiation of bosons,
     $g^{\mu\nu}$ in the normal Hamilton-Jacobi equation of bosons should be substituted by
     $g^{\mu\nu}+\lambda u^{\mu}u^{\nu}$, where $u$ is an etheric-like vector.
     Applying the inverse metric tensors of Vaidy-Bonner black hole and choosing
     a suitable $u$, one can obtain the dynamical equation for bosons in the Vaidy-Bonner spacetime.
      The solution process of  this new Hamilton-Jacobi equation is similar to the method in this paper,
      such as tortoise coordinate transformation, separation of variable. We will do a serial of researches to
     these problems in the future work.\\

This work was supported by the National Natural Science Foundation
of China (grant 11273020, 11847048), the Science Foundation of
Sichuan Science and Technology Department(grant 2018JY0502), the
Science Foundation of China West Normal University (grant 17YC050)
and the Natural Science Foundation of Shandong Province(grant
ZR2019MA059).





\end{document}